# On the level of background in underground muon measurements with plastic scintillator counters

V. I. Volchenko[1], G. V. Volchenko[1], E. V. Akhrameev[1], L. B. Bezrukov[1], I. M. Dzaparova[1], I. Sh. Davitashvili[1], T. Enqvist[2], H. Fynbo[4], Zh. Sh. Guliev[1], L. V. Inzhechik[1], A. O. Izmaylov[1], J. Joutsenvaara[2], M. M. Khabibullin[1], A. N. Khotjantsev[1], Yu. G. Kudenko[1], P. Kuusiniemi[2], B. K. Lubsandorzhiev[1*], O. V. Mineev[1], V. B. Petkov[1], R. V. Poleshuk[1], B. A. M. Shaibonov[1], J. Sarkamo[2], A. T. Shaykhiev[1], W. Trzaska[3], A. F. Yanin[1], N. V. Yershov[1]

[1]*Institute for Nuclear Research of the Russin Academy of Siences, Moscow Russia*

[2]*CUPP/Pyhäsalmi, University of Oulu, Oulu Finland*

[3]*Department of Physics, University of Jyväskylä, Jyväskylä Finland*

[4]*Department of Physics and Astronomy, University of Århus, Denmark*

∗*Corresponding author: postal address: pr-t 60th Anniversary of October, 7a, 117312 Moscow, Russia; phone: +7-495-1353161; fax: +7-495-1352268;*

E-mail: lubsand@pcbai10.inr.ruhep.ru, lubsand@pit.physik.uni-tuebingen.de

**Abstract**

In this short note we present results of background measurements carried out with a polystyrene based cast plastic $12.0\times12.0\times3.0$ cm$^3$ size scintillator counter with a wavelength shifting fibre and a multi-pixel Geiger mode avalanche photodiode readout in the Baksan underground laboratory at a depth of 200 meters of water equivalent. The total counting rate of the scintillator counter measured at this depth and at a threshold corresponding to ~0.37 of a minimum ionizing particle is approximately 1.3 Hz.

PACS: 29.40.Mc; 85.60.Dw; 95.85

Key words: Plastic scintillator, cosmic ray, muon, radioactivity background, multi-pixel avalanche photodiode.

Plastic scintillator counters are presently widely used in many underground experiments as scintillator muon-veto counters in neutrinoless double beta-decay experiments (e.g. GERDA [1]), dark matter search experiments (e.g. CRESST-II [2]) and cosmic ray muon experiments like EMMA [3, 4]. A scintillator counter is usually a plastic scintillator plate which is thick enough in order to have a good amplitude response to cosmic ray muons or minimum ionizing particles (MIP). In all such experiments the main problem is how to reach high efficiency for cosmic ray muon detection while at the same time provide high efficiency of background rejection. The background is mainly caused by radioactive decays in the rock surrounding the detector. The background spectrum lies in a lower energy domain in comparison to muon events but has a rather long tail overlapping with the muon spectrum. Thus one should set the energy threshold such that both muon detection efficiency and background rejection are at an acceptable level for the experiment. This can be carried out if there is a good separation between muon and background events. Energy deposition of a relativistic muon in plastic scintillator depends only on the scintillator thickness. One solution is to use thicker scintillator plates.

We have carried out a dedicated measurement of the background counting rate and muon detection efficiency at a depth of 200

meters of water equivalent (m.w.e.) in the Baksan underground laboratory using a small plastic scintillator counter manufactured for the cosmic ray muon experiment EMMA at Pyhäsalmi mine in Finland [3, 4].

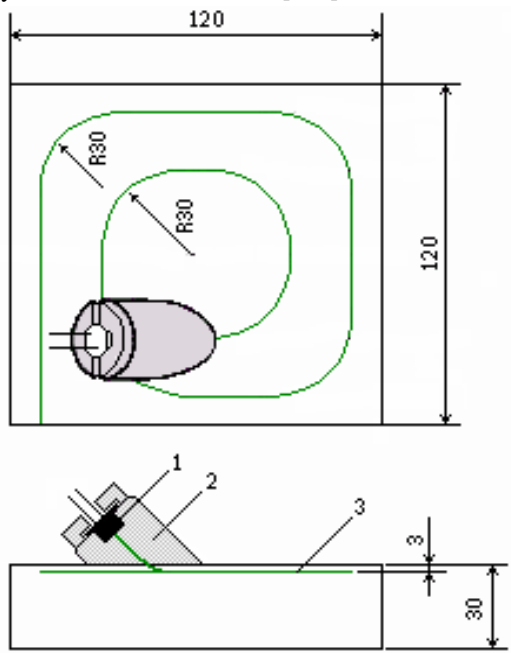

Fig.1. Schematic view of the scintillator counter. 1 – MRS APD; 2 – connector; 3 – WLS fibre.

A schematic view of the scintillator counter is shown in Fig. 1. The scintillator counter is a polystyrene based cast plastic scintillator of $12.0 \times 12.0 \times 3.0$ cm$^3$ with a 3 mm deep spiral groove cut into one side. A 0.5 m long Kuraray Y11 (200 ppm dopant) wavelength shifting (WLS) fibre of ~1 mm diameter has been placed in the groove and optically coupled with the scintillator bulk material using Bicron BC 600 optical cement. One of the fibre ends is polished and connected to a multi-pixel avalanche photodiode. The other end is also polished and mirrored.

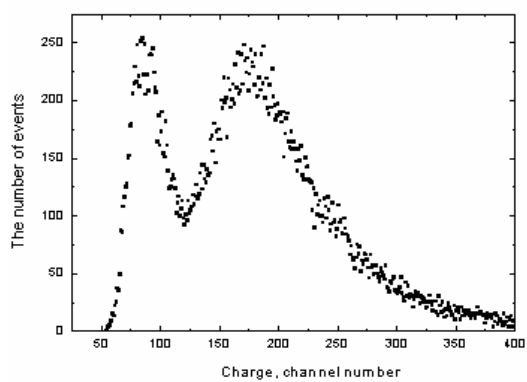

Fig.2. Charge spectrum of the events detected with the scintillator counter in the surface laboratory.

The multi-pixel avalanche photodiode with a metal-resistor-semiconductor layer structure operating in the limited Geiger mode (known as MRS APD) is produced by the CPTA Company, Moscow. The photodiode has a 1.1 mm$^2$ sensitive area with 556 pixels of $45 \times 45$ μm$^2$ each. We refer to [5-7] for more details on MRS APD. The operating voltage of the photodiode used in the measurements was 52.5V providing a photon detection efficiency (PDE) of ~20% at $\lambda$=515 nm at room temperature (22°C). The dark current counting rate of the photodiode was about 1.2 MHz above threshold of ~0.5 photoelectrons (pe). The MRS APD gain was ~$5 \times 10^5$.

The charge spectrum of events measured with the scintillator counter in the surface laboratory in a self-triggered mode at room temperature is shown in Fig. 2. The cut off in the left part of the spectrum is due to a discriminator threshold of ~23 pe. The total counting rate of the counter above this threshold is ~3.7 Hz. A peak produced by cosmic ray muons is clearly seen on the right and corresponds to ~56 pe. The peak-to-valley ratio of the spectrum is ~2.5. As illustrated in Fig. 2, the counter has a muon detection efficiency of almost 100%.

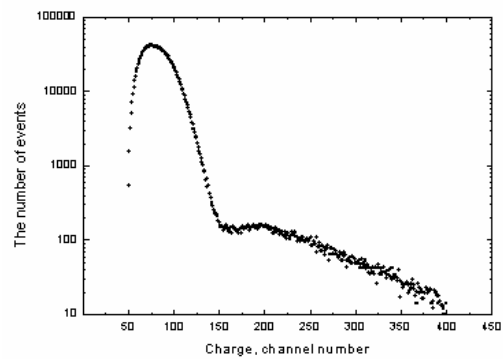

Fig.3. Charge spectrum of the events detected with the scintillator counter at the underground laboratory at the depth of 200 m.w.e..

The charge spectrum of events measured with the same scintillator counter in the underground laboratory at the depth of 200 m.w.e. is presented in Fig. 3. The ADC pedestal was subtracted in both spectra. The discriminator threshold was the same as in the measurements at the surface. The temperature in the underground laboratory was stabilized to a level of ~14.5±0.5 $^0$C throughout the measurements. The total number of counts in the spectrum is approximately $1.75 \times 10^6$ within the total time exposure of $1.39\times10^6$ s. The largest contribution to the spectrum is from background due to radioactivity in the surrounding rock. The muon peak in the spectrum corresponds to ~61 pe. The observed shift in the muon peak position in comparison to the spectrum registered in the surface laboratory is due to an increase of the photodiode gain and photon detection efficiency because of the temperature decrease [5, 6]. As can be seen in Fig. 3, the cosmic ray muons contribution to the total spectrum of the scintillator counter is negligibly small at such threshold but at the same time the muon detection efficiency is sufficiently high. Thus the total counting rate of the scintillator counter was ~1.3 Hz above the threshold of ~0.37 MIP and almost entirely due to radioactivity in the rock surrounding the underground laboratory.

The results of the background counting rate measurement with the scintillator counter presented in this paper will be useful for the EMMA experiment as well as for other currently running and being planned underground experiments.

The authors would like to thank Dr. V. Ch. Lubsandorzhieva and Dr. D. G. Middleton for careful reading of the manuscript and many valuable remarks and discussions.